\documentclass[conference]{IEEEtran}
\usepackage{amsmath,graphicx} 
\usepackage{amssymb}
\usepackage{algpseudocode}
\usepackage{cite}
\usepackage{amsfonts}
\usepackage{graphicx,subfigure}
\usepackage{fancyhdr}
\usepackage{cases}
\usepackage{extarrows}
\usepackage{algorithm}
\usepackage{multirow,tabularx}
\usepackage{mathtools}
\usepackage{xcolor}
\usepackage{epsfig}
\usepackage[english]{babel}

\IEEEoverridecommandlockouts

\begin{document}
\title{Energy Efficient Multi-User MISO Communication using Low Resolution Large Intelligent Surfaces}
\author{
\IEEEauthorblockN{Chongwen~Huang$^1$, George~C.~Alexandropoulos$^2$, Alessio Zappone$^3$, M\'{e}rouane Debbah$^{2,3}$, and Chau Yuen$^1$
\thanks{The work of C. Yuen was supported by the MIT-SUTD International design center and NSFC 61750110529 Grant, and that of C. Huang by the PHC Merlion PhD program. The work of M. Debbah was supported by H2020 MSCA IF BESMART, Grant 749336, and H2020-ERC PoC-CacheMire, Grant 727682; the former project also funded the work of A. Zappone.}
 }
\IEEEauthorblockA{
$^1$Singapore University of Technology and Design, 487372 Singapore\\
emails: chongwen\_huang@mymail.sutd.edu.sg, yuenchau@sutd.edu.sg\\
$^2$Mathematical and Algorithmic Sciences Lab, Huawei Technologies France SASU, 92100 Boulogne-Billancourt, France\\
emails: \{george.alexandropoulos, merouane.debbah\}@huawei.com\\
$^3$CentraleSup\'elec, Universit\'e  Paris-Saclay, 91192 Gif-sur-Yvette, France\\
email: alessio.zappone@unicas.it}}

\maketitle

\begin{abstract}
We consider a multi-user Multiple-Input Single-Output (MISO) communication system comprising of a multi-antenna base station communicating in the downlink simultaneously with multiple single-antenna mobile users. This communication is assumed to be assisted by a Large Intelligent Surface (LIS) that consists of many nearly passive antenna elements, whose parameters can be tuned according to desired objectives. The latest design advances on these surfaces suggest cheap elements effectively acting as low resolution (even $1$-bit resolution) phase shifters, whose joint configuration affects the electromagnetic behavior of the wireless propagation channel. In this paper, we investigate the suitability of LIS for green communications in terms of Energy Efficiency (EE), which is expressed as the number of bits per Joule. In particular, for the considered multi-user MISO system, we design the transmit powers per user and the values for the surface elements that jointly maximize the system's EE performance. Our representative simulation results show that LIS-assisted communication, even with nearly passive $1$-bit resolution antenna elements, provides significant EE gains compared to conventional relay-assisted communication.
\end{abstract}


\section{Introduction}\label{sec:intro}
The highly demanding data rate requirements for fifth Generation (5G) and beyond wireless networks, which are anticipated to connect over $50$ billions of wireless devices by 2020 \cite{EricssonWP} via dense deployments of multi-antenna base stations and access points \cite{mmwave_5G,Alexandropoulos_CB}, have raised serious concerns on their energy consumption footprint. To address the increasingly critical need for green and sustainable emerging and future networks \cite{5GNGMN,Zap2016,Green_com}, several energy efficient wireless solutions have been lately proposed \cite{ZapNow15,GEJSAC16}, ranging from renewable energy sources and energy efficient hardware to green resource allocation and transceiver signal processing techniques.

Among the recent transceiver hardware technologies \cite{Reconfigurable_arrays,alexandg_ESPARs,Ralf_LMA} with significant potential in reducing the energy consumption of wireless networks, while being theoretically capable of offering unprecedented massive Multiple-Input Multiple-Output (MIMO) gains \cite{scaling_up,phaseshifter_constraits,emil_setting,sha_hu} belong the Large Intelligent Surfaces (LIS). These surfaces are man made structures that can be electronically controlled with integrated electronics and wireless communication. A LIS usually comprises of a vast amount of small and nearly passive reflecting elements with reconfigurable parameters. Current implementations of LIS include conventional reflectarrays \cite{Reconfigurable_arrays,tan_indoor,tan_infocom2018}, liquid crystal metasurfaces \cite{Foo_LC}, or even software defined metamaterials intended for nanonetworks \cite{Liaskos_metasurface}. The LIS reflecting elements are usually very low cost and energy consumption units whose tuning may affect the electromagnetic behavior of the wireless propagation channel. Each of these units can effectively reflect a phase shifted version of the impinging electromagnetic field, hence, the combined configurations of all LIS elements may achieve certain communication objectives. Although LIS operation resembles that of a multi-antenna relay \cite{relay_model}, it is fundamentally different from relaying. LIS performs as a reconfigurable scatterer and does not require any dedicated energy source for either decoding, channel estimation, or transmission. In addition, intelligent surfaces can be easily placed into room and factory ceilings, buildings facades, and laptop cases, up to being integrated into human clothing.

The LIS parameters design for various communication objectives has been the focus of the recent research works \cite{chongwen2018,Subrt_control,tan_indoor,Subrt_control01,sha_hu,Reconfigurable_arrays,tan_infocom2018,Liaskos_metasurface}. The vast majority of the theoretical investigations have considered infinite resolution values for the LIS reflecting elements. In \cite{Subrt_control,Subrt_control01}, the role of LIS consisting of passive elements for improving indoor coverage was analyzed. A detailed analysis on the information transfer from multiple users to a LIS with active elements was carried out in \cite{sha_hu}. Very recently, \cite{tan_infocom2018} experimented on the incorporation of a smart reflectarray in a IEEE 802.11ad network operating in the unlicensed $60$GHz frequency band. In \cite{chongwen2018}, considering a LIS with infinite phase resolution passive elements, it was shown that higher spectral efficiencies can be achieved when LIS-assisted communication is a feasible option.

In this paper, we focus on a multi-user Multiple-Input Single-Output (MISO) communication system assisted by a LIS comprised of nearly passive reflecting elements with only low phase resolution tuning capabilities. We study the Energy Efficiency (EE) maximization problem and present an algorithm for the joint design of the transmit powers for the users and the phase values for the LIS elements. Our numerical results showcase that LIS with even $1$-bit phase resolution elements can provide significant EE performance gains compared to conventional relay-assisted communication.

\section{System Model}\label{sec:format}
In this section, we present the signal model for the considered LIS-assisted downlink multi-user MISO system, as well as our adopted model for the system total power consumption.
\begin{figure}
  \begin{center}
  \includegraphics[width=80mm]{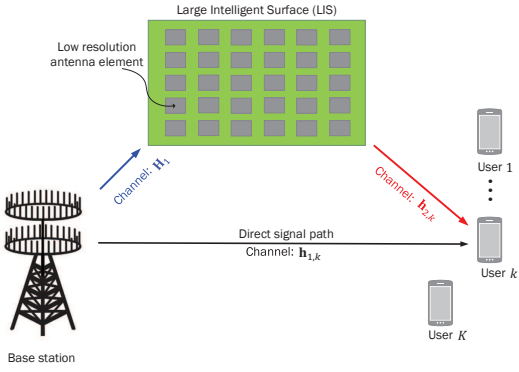}  \vspace{-2mm}
  \caption{The considered LIS-assisted multi-user MISO system comprising of a $M$-antenna base station simultaneously serving in the downlink $K$ single-antenna users. LIS is assumed to be attached to a surrounding building's facade, and the transmit signal propagates to the users both directly from the base station and via the LIS with reconfigurable behavior.}
  \label{fig:Estimation_Scheme} \vspace{-6mm}
  \end{center}
\end{figure}

\subsection{Signal Model}
We consider the wireless communication system illustrated in Fig$.$~\ref{fig:Estimation_Scheme}. This system consists of a Base Station (BS) equipped with $M$ antenna elements that wishes to convey information bearing signals simultaneously to $K$ single-antenna mobile users. This communication is assisted by a LIS attached to the facade of a building existing in the vicinity of the multi-user MISO system. The LIS is capable of reconfiguring the electromagnetic behavior of the wireless propagation channel according to desired objectives. It is comprised of $N\geq K$ cheap and nearly passive reflecting elements \cite{Reconfigurable_arrays}, which effectively act as low resolution phase shifters impacting the impinging information bearing electromagnetic field.

We denote by $\mathbf{h}_{1,k}\in\mathbb{C}^{1\times M}$ the direct channel between BS and the $k$-th mobile user, where $k=1,2,\ldots,K$. Similarly, $\mathbf{h}_{2,k}\in\mathbb{C}^{1\times N}$ represents the channel between LIS and the $k$-th mobile user, and we use the notation $\mathbf{H}_{1}\in \mathbb{C}^{N\times M}$ for the channel between BS and LIS. The entries inside all these matrices are modeled as Independent and Identically Distributed (IID) complex Gaussian random variables with variance depending on the pathloss of the respective wireless links. We also assume that there's no dependence or correlation between the elements of any pair of different channel matrices. It can be shown that the baseband representation of the received signal at the $k$-th mobile user is given by
\begin{equation}\label{model_01}
y_k = \left(\mathbf{h}_{2,k}\mathbf{\Phi}\mathbf{H}_{1}+\mathbf{h}_{1,k}\right)\mathbf{x}+n_k,
\end{equation}
where $\mathbf{x}\in \mathbb{C}^{M \times 1}$ denotes the transmitted signal from BS antenna elements comprising of the individually precoded signals for each of the $K$ users; $\mathbf{\Phi}\triangleq\mathrm{diag}[\phi_1,\phi_2,\ldots,\phi_N]$ is a diagonal matrix including the effective phase shifting values $\phi_n$ $\forall$$n=1,2,\ldots,N$ for all LIS elements; and $n_k\sim\mathcal{CN}(0, \sigma^2)$ models the zero mean complex Additive White Gaussian Noise (AWGN) with variance $\sigma^2$. The transmitted signal $\mathbf{x}$ can be expressed as $\mathbf{x}\triangleq\sum_{k=1}^{K}\sqrt{p}_{k}\mathbf{g}_{k}s_{k}$ with $p_{k}$, $s_{k}$, and $\mathbf{g}_{k}$ representing the transmit power, unit power complex valued information symbol chosen from a discrete constellation set, and precoding vector intended for $k$-th mobile user. In practice, signal $\mathbf{x}$ is subject to a transmit power constraint $P>0$ indicating the maximum allowable transmit power. To this end, it holds for the expectation of the transmit signal power that
\begin{equation}\label{model_03}
  E\{|\mathbf{x}|^2\}=\mathrm{tr}(\mathbf{P}\mathbf{G}^H\mathbf{G})\leq P,
\end{equation}
where $\mathrm{tr}(\cdot)$ is the trace operand, $\mathbf{G}\triangleq[\mathbf{g}_1,\mathbf{g}_2,\ldots,\mathbf{g}_K]\in\mathbb{C}^{M\times K}$, and $\mathbf{P}\triangleq\mathrm{diag}[p_1,p_2,\ldots,p_K]$. We finally consider the following expression for the effective phase shifting value (finite resolution) for the $n$-th element of the LIS:
\begin{equation}\label{model_05}
\phi_n \in \mathcal{F} \triangleq \left\{\exp\left(\frac{j2\pi m}{2^b}\right)\right\}_{m=0}^{2^b-1},
\end{equation}
where $j\triangleq\sqrt{-1}$ is the imaginary unit, $\mathcal{F}$ represents the set with the available phase shifting values, $m$ denotes the phase shifting index in \eqref{model_05}, and $b$ gives the phase resolution in number of bits. Clearly, the different number of phase shifting values per LIS element is $2^b$. This number determines the LIS hardware complexity and power consumption.

It easily follows from \eqref{model_01} that the Signal-to-Interference-plus-Noise Ratio (SINR) at the $k$-th user is expressed as
\begin{equation}\label{model_06}
  \gamma_k =\frac{p_k|(\mathbf{h}_{2,k}\mathbf{\Phi} \mathbf{H}_{1}+\mathbf{h}_{1,k})\mathbf{g}_k|^2}{\sum\limits_{i=1,i\neq k}^{K} p_i|(\mathbf{h}_{2,k}\mathbf{\Phi} \mathbf{H}_{1}+\mathbf{h}_{1,k})\mathbf{g}_i|^2+\sigma^2}.
\end{equation}
Hence, the achievable sum rate performance of the considered LIS-assisted downlink multi-user MISO system is obtained as
\begin{equation}\label{model_07}
  \mathcal{R} =\sum\limits_{k=1}^{K}\mathrm{log_2}(1+\gamma_k).
\end{equation}

\subsection{Total Power Consumption Model}
Let us consider the point-to-point communication link between BS and the $k$-th mobile user. Recall that $p_k$ represents the transmit power allocated to the signal intended for the $k$-th user. A realistic expression for the consumed power for this $k$-th wireless communication link reads as \cite{ZapNow15}
\begin{equation}\label{power_model1}
   \mathcal{P}_k =\mu p_k + P_c + P_{\rm LIS},
\end{equation}
where $\mu \triangleq \nu^{-1}$ with $\nu$ being the efficiency of the transmit power amplifier, while $P_c$ incorporates the power dissipated in all other circuit blocks of the transmitter and receiver to operate the communicating terminals. Finally, $P_{\rm LIS}$ denotes the LIS power consumption. We should remark that the two underlying assumptions in \eqref{power_model1} are: \textit{i}) the transmit amplifier operates in its linear region; and \textit{ii}) the circuit power $P_c$ does not depend on the communication rate $R$.  Both assumptions are met in typically wireless communication systems, which are operated so as to ensure that the amplifiers operate in the linear region of their transfer function, and in which the hardware-dissipated power is just a constant power offset.

The LIS power consumption depends on the type and the resolution of its reflecting elements that effectively perform phase shifting on the impinging signal. By assuming that each elements is actually a phase shifter, typical values of its consumed power are $15$, $45$, $60$, and $78$mW for $3$-, $4$-, $5$-, and $6$-bit resolution phase shifting \cite{phaseshifter_powerconsup}. In the latter power values, the power consumption of the low noise amplifier has also been included. Therefore, the power dissipated at an intelligent surface with $N$ identical reflecting elements can be written as:  \vspace{-2mm}
\begin{equation}\label{power_model2} 
   P_{\rm LIS} = N P_{n}(b), \vspace{-2mm}
\end{equation} 
where $P_{n}(b)$ denotes the power consumption of each phase shifter having $b$-bit resolution. Putting all above together (i$.$e$.$, \eqref{power_model1} and \eqref{power_model2}), the total amount of power needed to operate our LIS-assisted downlink multi-user MISO system is given by \vspace{-1.8mm}
\begin{equation}\label{power_model3} \vspace{-1mm}
   \mathcal{P}_{\rm total}=\sum_{k=1}^{K}\mu_k p_k + KP_{c} + NP_{n}(b).
\end{equation}

\section{Design Problem formulation}
We are interested in the joint design of the transmit powers for all users, included in $\mathbf{P}=\mathrm{diag}[p_1,p_2,\ldots,p_K]$, and the values for the LIS elements, appearing in the diagonal of $\mathbf{\Phi}=\mathrm{diag}[\phi_1,\phi_2,\ldots,\phi_N]$, that jointly maximize the EE performance of the considered LIS-assisted multi-user MISO system, while satisfying the individual Quality of Service (QoS) requirements of the $K$ mobile users. To make the targeted problem more tractable, we assume that all involved channels are perfectly known at BS that employs Zero-Forcing (ZF) transmission, which is known to be optimal in the high-SINR regime \cite{DE_ZFR}. Note that the practical estimation (either partial or explicit) of especially $\mathbf{h}_{2,k}$ $\forall$$k=1,2,\ldots,K$ and $\mathbf{H}_1$ is a difficult task that will require considering sophisticated methods (e$.$g$.$, \cite{Alexandropoulos_position, Alkhateeb_DNN}); this is left as future work.

The ZF precoding matrix to be substituted into \eqref{model_06} is given by $\mathbf{G}=(\mathbf{H}_{2}\mathbf{\Phi} \mathbf{H}_1+\mathbf{H})^{+}$, where $(\cdot)^{+}$ denotes pseudo-inversion, $\mathbf{H}\triangleq[\mathbf{h}_{1,1}^{\rm T},\mathbf{h}_{1,2}^{\rm T},\ldots,\mathbf{h}_{1,K}^{\rm T}]^{\rm T}\in\mathbb{C}^{K\times M}$, and $\mathbf{H}_2\triangleq[\mathbf{h}_{2,1}^{\rm T},\mathbf{h}_{2,2}^{\rm T},\ldots,\mathbf{h}_{2,K}^{\rm T}]^{\rm T}\in\mathbb{C}^{K\times N}$. Then, the EE performance, $\eta$, for our system measured in bits/Joule is defined as the ratio of the achievable sum rate over the total power consumption. It can, therefore, be computed using \eqref{model_07} for the case of ZF precoding and \eqref{power_model3} as $\eta\triangleq \mathcal{R}/\mathcal{P}_{\rm total}$. The considered EE maximization problem is finally expressed as follows:
\begin{subequations}\label{Prob:ResAllpower}
\begin{align}
&\displaystyle \max_{\mathbf{\Phi},\mathbf{P}}\;\frac{\sum_{k=1}^{K}\log_2\left(1+\frac{p_k}{\sigma^2}\right)}{\sum_{k=1}^{K}\mu_k p_k+KP_{c}+NP_{n}(b) }\label{Prob:aProb:ResAllpower}\\
&\;\text{s.t.}\;\log_2\left(1+\frac{p_k}{\sigma^2}\right)\geq R_{{\rm min},k}\;\forall k=1,2,\ldots,K,\label{Prob:bProb:ResAllpower}\\
&\;\quad\;\; \text{tr}((\mathbf{H}_{2}\mathbf{\Phi} \mathbf{H}_{1}+\mathbf{H})^{+}\mathbf{P}((\mathbf{H}_{2}\mathbf{\Phi} \mathbf{H}_{1}+\mathbf{H})^{+})^{\rm H})\leq P,\label{Prob:cProb:ResAllpower}\\
&\;\quad\;\; \phi_n\in\mathcal{F}=\{1,e^{j2^{1-b}\pi},\ldots,e^{j2\pi(2^b-1)/2^b}\},\;b = 1,2,\ldots \nonumber\\
&\;\quad\;\; \forall n=1,2,\ldots,N, \label{Prob:dProb:ResAllpower}
\end{align}
\end{subequations}
where $R_{{\rm min},k}$ denotes the individual QoS constraint of the $k$-th user. Recall that in the rate expressions included in the latter problem we have considered the impact of ZF precoding.

The optimization problem in \eqref{Prob:ResAllpower} is non-convex, and especially the optimization with respect to $\mathbf{\Phi}$
is challenging due to the integer nature of the phase shifting LIS elements. An exhaustive search approach would have the exponential complexity $\mathcal{O}(2^{bN}NKM)$, which is prohibitive when considering large surfaces. In the sequel, we develop a low complexity EE maximization algorithm for the practically interesting LIS case having few bits resolution phase shifting elements.

\section{Transmit Power Allocation and LIS Design}
Obtaining the jointly optimal LIS phase shifting matrix $\mathbf{\Phi}$ and the transmit power allocation matrix $\mathbf{P}$ solving the optimization problem \eqref{Prob:ResAllpower} is a cumbersome task, mainly due to the constraint \eqref{Prob:dProb:ResAllpower} . One convenient approach is to employ an alternating optimization approach \cite{J:alternating_minimization} to separately and iteratively solve for $\mathbf{P}$ and $\mathbf{\Phi}$. In doing so, we first assume a fixed $\mathbf{P}$ and solve for $\mathbf{\Phi}$ maximizing the EE objective function in \eqref{Prob:ResAllpower}. Then, keeping $\mathbf{\Phi}$ fixed, we find $\mathbf{P}$ optimizing EE performance. This procedure is repeated until reaching convergence of the objective function or the solutions to an acceptable level. In the following, we first design an algorithm for the case where the LIS elements have $1$-bit phase resolution. We then generalize that algorithmic approach to LIS cases having any finite resolution phase shifting value.

\subsection{LIS with $1$-bit Phase Resolution Elements}
Following \eqref{model_05} for this lowest phase resolution case of $b=1$ yields $\phi_n=\{1,-1\}$ $\forall$$n=1,2,\ldots,N$. In this practically interesting case, the optimization problem \eqref{Prob:ResAllpower} becomes
\begin{subequations}\label{Prob:onebit}
\begin{align}
&\displaystyle \max_{\mathbf{\Phi},\mathbf{P}}\;\frac{\sum_{k=1}^{K}\log_2\left(1+\frac{p_k}{\sigma^2}\right)}{\sum_{k=1}^{K}\mu_k p_k+KP_{c}+NP_{n}(1) }\label{Prob:aProb:onebit}\\
&\;\text{s.t.}\;\log_2\left(1+\frac{p_k}{\sigma^2}\right)\geq R_{{\rm min},k}\;\forall k=1,2,\ldots,K,\label{Prob:bProb:onebit}\\
&\;\quad\;\; \text{tr}((\mathbf{H}_{2}\mathbf{\Phi} \mathbf{H}_{1}+\mathbf{H})^{+}\mathbf{P}((\mathbf{H}_{2}\mathbf{\Phi} \mathbf{H}_{1}+\mathbf{H})^{+})^{\rm H})\leq P,\label{Prob:cProb:onebit}\\
&\;\quad\;\; \phi_n=\{1,-1\} \; \forall n=1,2,\ldots,N.  \; \label{Prob:dProb:onebit}
\end{align}
\end{subequations}
We next present the main alternating optimization steps for solving \eqref{Prob:onebit}. We first assume a fixed $\mathbf{P}$ and solve for $\mathbf{\Phi}$, and then keeping $\mathbf{\Phi}$ fixed, we find $\mathbf{P}$ maximizing \eqref{Prob:onebit}'s objective.

\subsubsection{Optimization with respect to $\mathbf{\Phi}$}
For a fixed transmit power allocation matrix $\mathbf{P}$, \eqref{Prob:onebit} becomes the feasibility test:
\begin{subequations}\label{Eq:ResAllProbPhi0}
\begin{align}
&\displaystyle \max_{\mathbf{\Phi}}\;1\\
&\;\text{s.t.}\;\text{tr}((\mathbf{H}_{2}\mathbf{\Phi} \mathbf{H}_{1}+\mathbf{H})^{+}\mathbf{P}((\mathbf{H}_{2}\mathbf{\Phi} \mathbf{H}_{1}+\mathbf{H})^{+})^{\rm H})\leq P, \label{Eq:bResAllProbPhi0}\\
&\;\quad\;\; \theta_{n}=\{0,\pi\} \;\forall n=1,2,\ldots,N,  \label{Eq:cResAllProbPhi0}
\end{align}
\end{subequations}
where we have expressed constraint \eqref{Prob:dProb:onebit} in terms of the phases of $\phi_n$'s; we have particularly defined $\phi_n\triangleq\exp(j\theta_{n})$. 

The challenge in solving \eqref{Eq:ResAllProbPhi0} lies in the fact that its objective is non-differentiable and that \eqref{Eq:cResAllProbPhi0} is a non-convex constraint. To proceed further, we observe that the LIS phase design problem of \eqref{Eq:ResAllProbPhi0} is feasible if and only if the solution of the following optimization problem:
\begin{subequations}\label{Eq:ResAllProbPhi12}
\begin{align}
&\displaystyle \min_{\mathbf{\Phi} }\text{tr}((\mathbf{H}_{2}\mathbf{\Phi} \mathbf{H}_{1}+\mathbf{H})^{+}\mathbf{P}((\mathbf{H}_{2}\mathbf{\Phi} \mathbf{H}_{1}+\mathbf{H})^{+})^{\rm H}) \label{Eq:aResAllProbPhi12}\\
&\;\text{s.t.}\; \theta_{n}=\{0,\pi\} \; \forall n=1,2,\ldots,N \label{Eq:bResAllProbPhi12}
\end{align}
\end{subequations}
is such that its objective can be made lower than the total transmit power constraint $P$. In order to solve \eqref{Eq:ResAllProbPhi12}, we proceed similar to \cite{phaseshifter_constraits}. Particularly, we relax the $\theta_n$ constraint for taking two discrete values to the weaker, but convex, constraint $0\leq\theta_n\leq2\pi$. Then, \eqref{Eq:ResAllProbPhi12} can be rewritten as
\begin{subequations}\label{Eq:ResAllProbPhi13}
\begin{align}
&\displaystyle \min_{\mathbf{\Phi} }\text{tr}((\mathbf{H}_{2}\mathbf{\Phi} \mathbf{H}_{1}+\mathbf{H})^{+}\mathbf{P}((\mathbf{H}_{2}\mathbf{\Phi} \mathbf{H}_{1}+\mathbf{H})^{+})^{\rm H}) \label{Eq:aResAllProbPhi13}\\
&\;\text{s.t.}\; 0 \leq \theta_n \leq 2\pi \;\forall n=1,2,\ldots,N. \label{Eq:bResAllProbPhi13}
\end{align}
\end{subequations}
The latter problem can be efficiently solved by numerical optimization methods, such as the interior-point and quasi-Newton methods. To this end, the function $\texttt{fmincon}$ in MATLAB can be leveraged. Note that constraint \eqref{Eq:bResAllProbPhi13} is linear, thus, the solution will be computed from $\texttt{fmincon}$ instantly. However, due to this constraint relaxation, the solution will yield $\theta_n$ values inside the interval $[0, 2\pi]$. To discretize the solution according to the phases' feasible set in \eqref{Eq:bResAllProbPhi12}, we set $\theta_n=0$ as the solution for \eqref{Eq:ResAllProbPhi12} when the solution for \eqref{Eq:ResAllProbPhi13} is such that $\frac{3\pi}{2}\leq \theta_n< 2\pi$ and $0 \leq \theta_n < \frac{\pi}{2}$. Similarly, we set $\theta_n=\pi$ when $\frac{\pi}{2} \leq\theta_n< \frac{3\pi}{2} $ from the solution of \eqref{Eq:ResAllProbPhi13}. It will be shown later on in the section with the performance evaluation results that the relaxation \eqref{Eq:bResAllProbPhi13} for the LIS phase values leads to a near-optimal $\mathbf{\Phi}$ when $N$ is moderately large.

\subsubsection{Optimization with respect to $\mathbf{P}$}\label{Sec:PowerAll}
For fixed LIS phase shifting matrix $\mathbf{\Phi}$, the optimization problem \eqref{Prob:ResAllpower} becomes
\begin{subequations}\label{Prob:fixedthetaEE}
\begin{align}
&\displaystyle \max_{\mathbf{P}}\;\frac{\sum_{k=1}^{K}\log_2\left(1+\frac{p_k}{\sigma^2}\right)}{\sum_{k=1}^{K}\mu_kp_k+KP_{c}+NP_n(1)}\label{Prob:afixedthetaEE}\\
&\;\text {s.t. } \;p_{k}\geq \sigma^{2}(2^{R_{{\rm min},k}}-1)\; \forall k=1,2,\ldots,K, \label{Prob:bfixedthetaEE}\\
&\;\quad\;\; \text{tr}((\mathbf{H}_{2}\mathbf{\Phi} \mathbf{H}_{1}+\mathbf{H})^{+}\mathbf{P}((\mathbf{H}_{2}\mathbf{\Phi} \mathbf{H}_{1}+\mathbf{H})^{+})^{\rm H})\leq P.\label{Prob:cfixedthetaEE}
\end{align}
\end{subequations}
It can be seen that the objective of the latter optimization problem is concave in $\mathbf{P}$ for fixed $\mathbf{\Phi}$. Moreover, both constraints \eqref{Prob:bfixedthetaEE} and \eqref{Prob:cfixedthetaEE} are convex with respect to $\mathbf{P}$. Hence, problem \eqref{Prob:fixedthetaEE} can be globally solved with limited complexity using Dinkelbach's method \cite{Dinkelbach}, as presented in Algorithm~\ref{Dinkelbach}. In this algorithm, set $\mathcal{B}\triangleq\{\mathbf{P}=\mathrm{diag}[p_1,p_2,\ldots,p_K]: \eqref{Prob:bfixedthetaEE}\wedge\eqref{Prob:cfixedthetaEE}\}$ and $\mathbf{P}^{*}_{i}\triangleq\mathrm{diag}[p^*_{1,i},p^*_{2,i},\ldots,p^*_{K,i}]$ denotes the transmit power allocation solution in Step 3 at each $i$-th (with $i=1,2,\ldots$) algorithmic iteration.
\begin{algorithm}[!t]
\caption{Dinkelbach's Method}\label{Dinkelbach}
\begin{algorithmic}[1]
\State \textbf{Initialization:} $\epsilon>0$ and $\lambda_0=0$.
\For{$i=1,2,\ldots$}
\State Solve the concave maximization:
\Statex \hspace{0.54cm}$\mathbf{P}^{*}_{i}=\arg\max\limits_{\mathcal{B}}\;\sum_{k=1}^{K}\log_2(1+\frac{p_k}{\sigma^2})$
\Statex \hspace{1.42cm}$-\lambda_{i-1}(\sum_{k=1}^{K}\mu_kp_k+KP_c+NP_n(1))$.
\State Set $\lambda_{i}=\frac{\sum_{k=1}^{K}\log_2\left(1+\frac{p^{*}_{k,i}}{\sigma^2}\right)}{\sum_{k=1}^{K}\mu_kp^*_{k,i}+KP_{c}+NP_n(1)}$.
    \If{$|\lambda_{i}-\lambda_{i-1}|<\epsilon$}
			\State \textbf{Output:} $\mathbf{P}^{*}_{i}$.
		\EndIf
\EndFor
\end{algorithmic}
\end{algorithm}

\subsubsection{Proposed EE Maximization Algorithm}\label{Sec:PowerAll}
The proposed EE maximization algorithm for our LIS-assisted multi-user MISO system for the case where the LIS is comprised of $1$-bit phase resolution reflecting elements is summarized in Algorithm~\ref{EEmax}. As shown, the presented solutions of \eqref{Eq:ResAllProbPhi13} for $\mathbf{\Phi}$ and \eqref{Prob:fixedthetaEE} for $\mathbf{P}$ are alternatively and iteratively deployed till reaching convergence of their solutions between consecutive runs (particularly, the squared norm of their difference) that is smaller than a small $\epsilon>0$. In this algorithm, subscript $\ell$ (with $\ell=1,2,\ldots$) in $\mathbf{\Phi}$ and $\mathbf{P}$ indicates their values at the $\ell$-th algorithmic iteration. Moreover, $\mathbf{I}_K$ and $\mathbf{0}_K$ denote the $K\times K$ identity and zeros matrices, respectively.
\begin{algorithm}[!t]
\caption{EE Maximization with $1$-bit Resolution LIS}\label{EEmax}
\begin{algorithmic}[1]
\State \textbf{Input:} $P$, $\sigma^2$, $\{R_{{\rm min},k}\}_{k=1}^{K}$, $\mathbf{H}_{1}$, and $\mathbf{H}_{2}$.
\State \textbf{Initialization:} $\mathbf{P}_{0}=\frac{P}{K}\mathbf{I}_K$, $\mathbf{\Phi}_{0}=\mathbf{0}_K$, and $\epsilon>0$.
\For{$\ell=1,2,\ldots$}
\State Find $\mathbf{\Phi}_\ell$ solving problem \eqref{Eq:ResAllProbPhi13} for the fixed $\mathbf{P}_{\ell-1}$.
\If{\eqref{Eq:bResAllProbPhi0} holds true for $\mathbf{\Phi}_\ell$}
 \State Find $\mathbf{P}_\ell$ solving problem \eqref{Prob:fixedthetaEE} using Algorithm~\ref{Dinkelbach}
 \Statex \hspace{1.067cm}for the fixed $\mathbf{\Phi}_\ell$.
\Else
 \State Break and declare infeasibility.
\EndIf
\If{$\|\mathbf{\Phi}_{\ell}-\mathbf{\Phi}_{\ell-1}\|^2<\epsilon$ and $\|\mathbf{P}_{\ell}-\mathbf{P}_{\ell-1}\|^2<\epsilon$}
 \State \textbf{Output:} $\mathbf{\Phi}_\ell$ and $\mathbf{P}_\ell$.
\EndIf
\EndFor
\end{algorithmic}
\end{algorithm}

\subsection{LIS with Finite Phase Resolution Elements}
We generalize the previously presented algorithmic solution for the LIS case of $1$-bit phase resolution to any finite resolution phase shifting value. To illustrate our generalization, let's consider first the $2$-bit resolution case where $\forall$$n=1,2,\ldots,N$: $\phi_n=\{1,j,-1,-j\}$ and equivalently $\theta_{n}=\{0,\frac{\pi}{2},\pi,\frac{3\pi}{2}\}$. In this case, the optimization over $\mathbf{\Phi}$ for fixed $\mathbf{P}$ is expressed as\vspace{-0.4cm}
\begin{subequations}\label{Eq:twobit}
\begin{align}
&\displaystyle \min_{\mathbf{\Phi} }\text{tr}((\mathbf{H}_{2}\mathbf{\Phi} \mathbf{H}_{1}+\mathbf{H})^{+}\mathbf{P}((\mathbf{H}_{2}\mathbf{\Phi} \mathbf{H}_{1}+\mathbf{H})^{+})^{\rm H}) \label{Eq:atwobit}\\
&\;\text{s.t.}\; \theta_{n}=\{0,\frac{\pi}{2},\pi,\frac{3\pi}{2}\} \; \forall n=1,2,\ldots,N. \label{Eq:btwobit}
\end{align}
\end{subequations}
Similar to the solution of \eqref{Eq:ResAllProbPhi12}, we relax constraint \eqref{Eq:btwobit} to $0\leq\theta_n\leq2\pi$. Then, we solve the relaxed problem using numerical optimization methods, as we treated problem \eqref{Eq:ResAllProbPhi13}. Finally, we discretize the solutions of the relaxed problem as: $\theta_n=0$ when the relaxed problem was solved with $0\leq\theta_n<\frac{\pi}{4}$ and $\frac{7\pi}{4}\leq\theta_n<2\pi$, $\theta_n =\frac{\pi}{2}$ for the interval $\frac{\pi}{4}\leq\theta_n<\frac{3\pi}{4}$, $\theta_n =\pi$ for $\frac{3\pi}{4}\leq \theta_n<\frac{5\pi}{4}$, and $\theta_n =\frac{3\pi}{2}$ when otherwise. Replacing this specific procedure within Algorithm~\ref{EEmax} extends the algorithm to the LIS case with $2$-bit phase resolution. In a similar way, we may use Algorithm~\ref{EEmax} to general LIS cases with any finite resolution value for their reflecting elements.

\section{Performance Evaluation Results}
\begin{figure}
  \begin{center}
  \includegraphics[width=75mm]{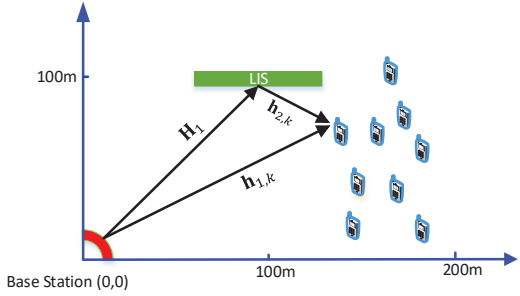}  \vspace{-2mm}
  \caption{The simulated LIS-assisted $K$-user MISO communication scenario comprising of a $M$-antenna base station and a $N$-element intelligent surface.}
  \label{fig:deploy} \vspace{-7mm}
  \end{center}
\end{figure}

We consider the LIS-assisted $K$-user MISO communication system scenario illustrated in Fig$.$~\ref{fig:deploy}. The multiple single-antenna mobile users are assumed randomly and uniformly placed in the $100m^2$ half right-hand side rectangular in between the channel realizations. For all average performance evaluation results, we have averaged over $500$ IID channel matrix realizations. In Fig$.$~\ref{fig:deploy}, BS is located in the origin point $(0,0)$ and the LIS central element is placed at the point $(100m,100m)$. Both structures are assumed to have their antenna elements placed uniformly on a rectangular surface (uniform planar arrays). To compute the total power consumption in a realistic way, we have used the propagation loss characterization described in \cite{Alessio2015EE,emil_setting}. For the hardware dissipation power model, we have set $P_{c}=100$dBm and $\mu_k=1.1$ $\forall$$k=1,2,\ldots,K$. The values $P_n(1)=5$dBm, $P_n(2)=15$dBm, and $P_n(+\infty)=45$dBm for the per LIS element power consumption for the cases of $1$-, $2$-, and infinite-bit resolution have been used, respectively. We have compared the LIS-assisted performance with that using an amplify-and-forward multi-antenna relay. Particularly, we have considered a relay according to \cite{relay_model} having no precoding capabilities (this will require in general explicit channel estimation), negligible reception noise (ideal case), the amplification gain factor $\alpha=0.3$, and transmit power of $60$dBm. In the performance plots that follow we use the transmit Signal Noise Ratio (SNR) parameter, defined as ${\rm SNR} = P/\sigma^2$, and we have run both Algorithms~\ref{Dinkelbach} and~\ref{EEmax} with the accuracy setting $\epsilon=0.01$.

\begin{figure}
  \begin{center}
  \includegraphics[width=75mm]{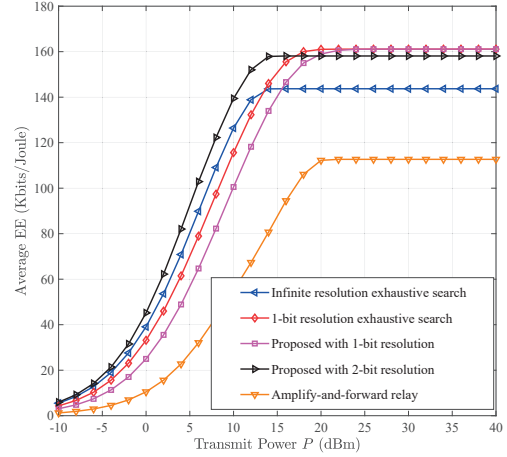}  \vspace{-2mm}
 \caption{Energy efficiency maximization vs the total BS transmit power $P$ for $K=16$, $M=12$, $N=32$, and $R_{{\rm min},k}=0$ bps/Hz $\forall$$k=1,2,\ldots,16$.}
  \label{fig:EEvsP} \vspace{-7mm}
  \end{center}
\end{figure}
Figure~\ref{fig:EEvsP} depicts EE versus $P$ in dBm for $K=16$, $M=12$, $N=32$, and the minimum QoS requirements $R_{{\rm min},k}=0$ bps/Hz $\forall$$k=1,2,\ldots,16$. The optimization problem \eqref{Prob:onebit} for the LIS case with $1$-bit phase resolution has been solved with two different ways: via exhaustive search and by using Algorithm~\ref{EEmax}. The exhaustive search approach has an exponential complexity, which is prohibitive for large $N$, but has been considered in this figure only for benchmarking purposes. We have also plotted EE results using the latter approaches for the LIS cases with $2$-bit and infinite phase resolution, as well as for an amplify-and-forward relay. In the latter case, we have solved a similar problem to \eqref{Prob:onebit}, where $\mathbf{\Phi}$ has been removed, using Algorithm~\ref{Dinkelbach}. In this relay-assisted case, only the transmit power allocation $\mathbf{P}$ has been designed. As shown in Fig$.$~\ref{fig:EEvsP}, EE for all schemes increases with increasing $P$ till a value above which it remains constant. This $P$ value is around $12.5$dBm for the $2$-bit and infinite resolution cases, and around $20$dBm for the $1$-bit resolution case. The latter two low resolution cases result in the highest EE, which is up to $45\%$ higher than that of relay-assisted system for $P>15$dBm. Evidently, up to $30\%$ EE improvement compared to the relay case can be also achieved with infinite resolution LIS elements.
\begin{figure}
  \begin{center}
  \includegraphics[width=75mm]{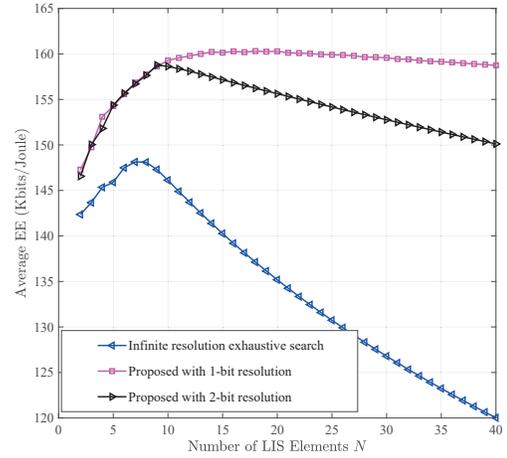}  \vspace{-2mm}
  \caption{Energy efficiency maximization vs the number of LIS elements $N$ for $K=8$, $M=6$, and $R_{{\rm min},k}=0$ bps/Hz $\forall$$k=1,2,\ldots,8$.}
  \label{fig:EEvsN} \vspace{-7mm}
  \end{center}
\end{figure}
The EE performance as a function of $N$ is plotted in Fig$.$~\ref{fig:EEvsN} for $K=8$, $M=6$, and $R_{{\rm min},k}=0$ bps/Hz $\forall$$k=1,2,\ldots,8$. It can be seen that for the few bits resolution values, both schemes exhibit the same trend: EE increases as $N$ increases up to a certain number. Indeed, increasing $N$ results in intelligent surfaces attaining their corresponding EE maximization. However, from a value of $N$ and on EE degrades. This happens because for large $N$ the LIS power consumption impacts EE in a negative way. Interestingly, the lower the resolution of the LIS elements, the slower is the EE degradation. Figure~\ref{fig:EEvsN} also indicates that there exists an optimal number of reflecting elements that maximizes EE for a considered LIS-assisted MISO communication system.

\begin{figure}
  \begin{center}
  \includegraphics[width=78mm]{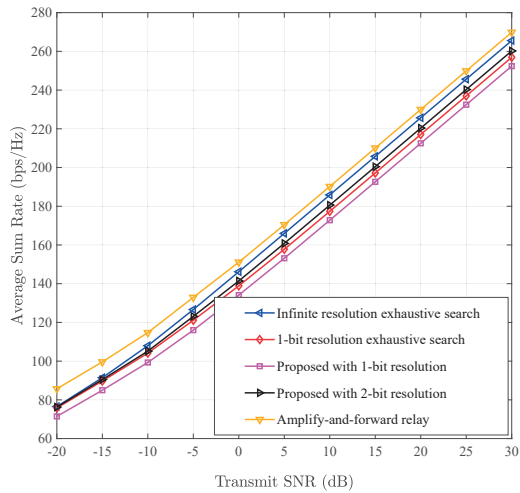}  \vspace{-2mm}
  \caption{Achievable sum rate vs the transmit ${\rm SNR}$ for $K=16$, $M=12$, $N=32$, and $R_{{\rm min},k}=\log_2(1+\frac{\mathrm{SNR}}{2K})$ bps/Hz $\forall$$k=1,2,\ldots,16$.}
  \label{fig:ombing} \vspace{-7mm}
  \end{center}
\end{figure}
In Fig$.$~\ref{fig:ombing}, we compare the achievable sum rate versus ${\rm SNR}$ in dB for the same setting of parameters with Fig$.$~\ref{fig:EEvsP}, except from the individual QoS constraints, which have been here set as $R_{{\rm min},k}=\log_2(1+\frac{\mathrm{SNR}}{2K})$ bps/Hz $\forall$$k=1,2,\ldots,16$. We have used the transmit power allocation designed in \cite{chongwen2018} for the sum rate maximization with relays. As depicted, relay-assisted communication achieves up to $20$bps/Hz more sum rate than the LIS-assisted one with $1$-bit resolution elements, as designed from Algorithm~\ref{EEmax}. For the case of infinite resolution elements, this performance gap reduces to less than $5$bps/Hz for large ${\rm SNR}$ values. It also evident that the larger the LIS resolution is, the larger is the sum rate that can be achieved. Note, however, that the $2$-bit phase resolution LIS case performs quite close to the infinite resolution one requiring much lower power consumption.

\section{Conclusion}\label{sec:prior}
In this paper, we have considered downlink multi-user MISO communication systems assisted by reconfigurable large surfaces. The surfaces are comprised of many nearly passive reflecting elements with mainly low phase resolution tuning capabilities. Intending at investigating the LIS suitability for green communications, we focused on the EE maximization problem and presented an algorithmic solution for the joint design of the transmit powers per mobile user and the phase values for the LIS elements. Our selected simulation results showcased that LIS with even $1$-bit phase resolution reflecting elements can increase the system's EE by more than $40\%$ compared to conventional amplify-and-forward relaying systems.

\bibliographystyle{IEEEtran}
\bibliography{IEEEabrv,references}

\end{document}